\documentstyle[osa,twocolumn]{revtex}

\begin{document}

\title{Turbulent spectrum of the earth's ozone field}
\author{L. Sirovich, R. Everson and D.Manin}
\address{The Rockefeller University, New York City, 10021}
\date{November 3, 1994}
\maketitle

\begin{abstract}
The Total Ozone Mapping Spectrometer (TOMS) database is subjected to an
analysis in terms of the Karhunen-Loeve (KL) empirical eigenfunctions. The
concentration variance spectrum is transformed into a wavenumber spectrum, $%
E_c(k)$. In terms of wavenumber $E_c(k)$ is shown to be $O(k^{-2/3})$ in the
inverse cascade regime, $O(k^{-2})$ in the enstrophy cascade regime with the
spectral {\it knee} at the wavenumber of barotropic instability.The spectrum
is related to known geophysical phenomena and shown to be consistent with
physical dimensional reasoning for the problem. The appropriate Reynolds
number for the phenomena is $Re\approx 10^{10}$.
\end{abstract}


\narrowtext
\vspace*{0.25in}
Atmospheric mixing is effected at horizontal scales which are large compared
with the scale height $\left( \approx 10{\it km}\right) $, which with
inhibition of vertical motion by planetary rotation and stable
stratification contributes to the two dimensional picture of atmospheric
activity.\cite{Ped}$^{\text{,}}$\cite{Phil} In addition to its essential
role in meteorology, current interest in mixing is enhanced by its role in
regard to the behavior of the Antarctica ozone {\it hole} \cite{Ssk} and to
the lack of such an {\it effect} in the Arctic. \cite{Petal}

Our investigation is based on satellite records of the earth's ozone fields.
We have analyzed fifteen years of daily ozone fields of the TOMS (Total
Ozone Mapping Spectrometer) database.\cite{Toms} Due to technical and
natural causes each daily record contains gaps in the form of missing
pixels. A typical snapshot appears in Figure 1. The dark regions represent
areas of missing pixels caused partly by technical failure and in part due
to polar night (measurements are based on reflected light). Each record is
{\it stitched} together from sixteen separate records obtained from
south-north synchronous orbits that are taken in a twenty four hour period
from the satellite, Nimbus.

The availability of such a large data set recommends statistical analyses,
and we focus on spectral properties of the ozone field. Although ozone
production (equatorial regions) and depletion (polar regions) result from
complex chemical reactions \cite{Chem} these represent relatively weak
sources and sinks and we follow common practice and regard ozone as a
passive scalar. The variance spectrum of a scalar contaminant in turbulent
flows has been recently reviewed by Sreenivasan. \cite{Sre} Functional
estimates for the concentration variance spectrum (in homogeneous isotropic
turbulence) follow from dimensional arguments based on those first given by
Kolmogorov leading to the famous $E_K\left( k\right) =K\epsilon
^{2/3}k^{-5/3}$ energy spectrum for the inertial range.\cite{Ko}

Both Obukhov and Corrsin\cite{ObCor} show that an inertial range can exist,
in particular the variance per wavenumber of concentration, $c$, denoted by $%
E_c(k),$ has the form,
\begin{equation}
E_c(k)=C\chi \epsilon ^{-1/3}k^{-5/3},  \label{one}
\end{equation}
where $C$ is a dimensionless constant, $\epsilon $ is the usual turbulent
energy transport rate, and $\chi =<\kappa (\nabla c)^2>$ is the appropriate
{\it dissipation} rate. Thus the concentration spectrum appears tied to the
corresponding velocity spectrum. A review of experimental observations\cite
{Sre} shows departures from the universal form (\ref{one}), except perhaps
at very high Reynolds numbers. For relatively small diffusional effects
Batchelor \cite{Bat} has shown that $E_c=O\left( k^{-1}\right) .$ This has
been shown to hold under less restrictive hypothesis.\cite{Kr1}$^{\text{,}}$%
\cite{Sig} Except for a recent simulation \cite{Sig} confirmation of this
result has been elusive. Predictions of anomolous scalings have also
appeared.\cite{Anom}

Arguments leading to the above spectra are unaltered when applied to two
dimensional turbulence. However, the interpretation of the cascade of energy
represented by the spectrum $E_K\left( k\right) $ requires some additional
remarks. Both energy $E$, and enstrophy $\Omega =$ $\int \left( \nabla
\wedge {\bf u}\right) ^2d{\bf x}$, are inviscid invariants in two
dimensions. As a result of this Kraichnan and Batchelor\cite{KrBa} have
shown that the Kolmogorov spectrum, $E_K$, represents an inverse cascade of
energy from smaller to larger scales, and that there also exists a second
cascade from small $k$ to large $k$ given by the enstrophy $\Omega
(k)=C_o\chi _ok^{-1}$(with log correction) and hence an energy spectrum $%
E(k)\propto k^{-3}$ , where $C_o$ is a dimensionless constant and $\chi
_0=\nu \overline{(\nabla \omega )^2}$ (also see\cite{Fal1}). Support for $%
E=O(k^{-3})$ comes from many direct simulations.\cite{Comp} However, recent
very large scale simulations show substantial divergence in the $O(k^{-5/3})$
inverse cascade range.\cite{Bo} Observational data from the atmosphere is
not definitive and although a power law energy spectrum is indicated in the
enstrophy range the exponent appears to lie between $-2\;$
and $-3$.\cite{Obs}
In particular, Schoeberl and Bacmeister\cite{Obs} suggest that the
exponent is $-2$ down to scales in the $10$ kilometer range!

A difficulty in interpreting these results for $E_c$ already appears.
Focusing on $k$ large, it might be supposed in analogy with three
dimensions, that $E_c$ $\left( k\right) =O\left( k^{-3}\right) $ i.e., it
should follow the energy spectrum. On the other hand, the vorticity (a
scalar) formally satisfies the same convection equation as does a passive
scalar, and from this one might suppose that $E_c=O\left( k^{-1}\right) $,
the Batchelor spectrum. As will be seen shortly, neither of these holds for $%
E_c$ in the atmosphere.

Other possible scalings for $E_c$ have appeared in the literature. For quasi
two-dimensional turbulence Falkovich and Medvedev \cite{Fal2} find $%
E=O\left( k^{-7/3}\right) $ for large k. Saffman, \cite{Saf} in considering
the Burgers equation, \cite{Bur} observed that its solutions are nearly
piecewise discontinuous which leads to $E_c=O\left( k^{-2}\right) $.
Pierrehumbert using concepts from chaotic mixing has obtained a variety of
scalings for $E_{c\text{ }}$from both mathematical and physical models.\cite
{Pie}

Satellite images (see Fig.1) are clearly inhomogeneous and a transformation
to wavenumber concepts is required. As will be seen the Karhunen-Loeve(KL)
procedure\cite{KL} is ideally suited for this purpose. In particular, the
snapshot method\cite{Qam}$\;$considerably reduces the needed computation
effort. However, the presence of gappy data required modification of the
methodology.\cite{Gap} This, as well as an extensive analysis of the
results, appears in Manin et al\cite{Man}$\;$ and a mathematical treatment
is also given elsewhere.\cite{Prep}

To connect the usual wavenumber spectrum with that obtained from the
empirical eigenfunctions we recall an earlier discussion.\cite{KnSi} The
concentration fluctuation of ozone is denoted by $c({\bf x},t)$. For
purposes of later dimensional reasoning we write the dimensions of $c$ as $%
\dim [c]=m/l^2$, where $m$ refers to molecules (of ozone) per area since the
data are two dimensional. The mean variance in the homogeneous case is given
by
\begin{equation}
\overline{c^2}=\frac 1A\int c^2d{\bf x}=\int {\cal E}_c({\bf k})d{\bf k}%
=\int E_c(k)dk.  \label{two}
\end{equation}
Thus $\dim [{\cal E}_c]=m^2l^{-2}$ and $\dim [E_c]=m^2l^{-3}$. To treat the
inhomogeneous case corresponding to the data we consider the correlation
\begin{equation}
K_c({\bf x},{\bf y})=<c({\bf x})c({\bf y})>  \label{thr}
\end{equation}
which from KL can be written in spectral form,

\begin{equation}
K_c=\sum_n\lambda _n\psi _n({\bf x})\psi _n({\bf y})  \label{fou}
\end{equation}
where $\{\psi _n\}$ are the eigenfunctions of the operator $K_c$. The total
variance is given by
\begin{equation}
<\int c^2({\bf x})d{\bf x}>=TrK=\sum_n\lambda _n.  \label{fiv}
\end{equation}
An eigenvalue $\lambda _n$ represents the average variance allocated to the
projection of $c$ onto $\psi _n$. The summation (\ref{fiv}) is the natural
generalization of (\ref{two}) to the inhomogeneous case. Each $\lambda _n$
represents the variance in a state, thus generalizing ${\cal E}_c({\bf k})$
and has the same dimensions,
\begin{equation}
\dim [\lambda ]=\dim [{\cal E}]=m^2l^{-2}.  \label{six}
\end{equation}

In Figure 2 we display in doubly logarithmic form $\lambda _n$ versus index $%
n$. As is seen the variance spectrum falls, to good approximation, on two
different power laws

\begin{equation}
\lambda _n\propto \left\{
\begin{array}{cc}
n^{\alpha _i};\;n<35,~~\alpha _i=.85\pm .035 &  \\
n^{\alpha _o};\;n>50,~~\alpha _o=-1.56\pm .022 &
\end{array}
\right.  \label{sev}
\end{equation}
The error bounds appearing in (\ref{sev}) and are based only on the least
squares fit to the data, and not on the methods used in arriving at the
spectrum which appears in Figure 2. The region of the {\it knee,} $35\leq
n\leq 50$, will be discussed below.

In order to relate $n$ to $k$ we observe that in the homogeneous case modes
carrying variances larger than those at $k$ can be counted in number, N, as
\begin{equation}
N\propto k^2,\;{\rm whence}\;k\propto N^{1/2}.  \label{eig}
\end{equation}

Before continuing this reasoning it is instructive to verify the accuracy
and content of these relations. For this purpose we employ the inverse
relation between wavenumber and length scale. Thus (\ref{eig}) implies that
the length scale, $L_n$, of the $n^{{\rm th}}$ eigenfunction bears the
following dependence on the index
\begin{equation}
L_n\propto n^{-1/2}.  \label{nin}
\end{equation}
An informal perusal of the eigenfunctions themselves supports this
relationship between characteristic length and index. To quantify this we
have computed the correlation length of each eigenfunction\cite{Man} and the
result is plotted in Figure 3. It is clear from this figure that (\ref{nin})
provides an excellent fit to the data in the two asymptotic regimes. The
region of the {\it knee} is the only anomaly and it appears as a plateau in
the figure and corresponds to just one scale.
\begin{equation}
2\pi /k_{*}=L_{*}\approx 4000\;{\rm km}.  \label{ten}
\end{equation}

It is generally stated in the geophysical literature\cite{Hou}$^{\text{,}}$%
\cite{Ped} that the baroclinic instability gives rise to a pattern. of
wavenumber roughly seven. I.e., the unstable pattern is made up of
approximately seven pairs of cyclonic/anti-cyclonic motions. With some
indulgence on the part of the viewer, eigenfunction $\psi _{44}$ shown in
Figure 4 appears to have this property. Roughly speaking, each of the
eigenfunctions in the range $35\leq n\leq 50$ shows this spatial
arrangement. To explain the plateau in Figure 3 we suggest an analogy with
the von Karman vortex trail. In that case a period seven disturbance
requires fourteen independent modes for its description.\cite{S85} In view
of the nature of the results this would appear to be a reasonable
explanation.

The energetics of the atmosphere has its origin in solar heating. However,
dynamical activity introduces $L_{*}$ as the lengthscale at which mechanical
energy is supplied, and is thus the significant lengthscale of the problem.
If ${\epsilon }$ is the energy transport rate which characterizes the
inverse energy cascade, then
\begin{equation}
\chi _0=k_{*}^2{\epsilon }  \label{ele}
\end{equation}
characterizes the enstrophy cascade to higher wavenumbers. Using estimates
for the physical parameters of the atmosphere and $L_{*}$ from (\ref{ten})
we obtain the Reynolds number, $R\approx 10^{10}$.

We now return to the implications of the power laws for $\lambda _n$, (\ref
{sev}) to the wavenumber spectrum. In keeping with customary practice we
consider the variance per wavenumber $E_c(k)=k{\cal E}_c(k)$. It follows
from (\ref{sev}) and (\ref{eig}) that
\begin{equation}
E_c(k)\propto \left\{
\begin{array}{ll}
k^{-2/3}, & k<k_{*} \\
k^{-2}, & k>k_{*}
\end{array}
\right.   \label{twe}
\end{equation}
The more precise exponents are entered for suggestive reasons. The high
wavenumber exponent lies slightly outside the error bound (However, it
should be noted that Kraichnan \cite{KrBa} actually estimated the enstrophy
fall-off as $1/k^3ln^{1/3}k$, see also \cite{Fal1} ). With the exception of
the Saffman-Burgers spectrum, \cite{Saf} theoretical predictions lie outside
the above range.

In view of the relationship (\ref{twe}), we now consider the consequences of
elementary dimensional analysis. A straightforward argument yields,
\begin{equation}
E_c(k)=\chi {\epsilon }^{-1/3}k_{*}^{-5/3}f(k/k_{*})  \label{thi}
\end{equation}
Equation (twe ) implies
\begin{equation}
f(k/k_{*})\sim \left\{
\begin{array}{ll}
c_i(k/k_{*})^{-2/3}, & k/k_{*}\ll 1 \\
c_o(k_{*}/k)^2, & k/k_{*}\gg 1
\end{array}
\right.   \label{four}
\end{equation}
where $c_i$ and $c_o$ are dimensionless constants. In these terms
\begin{equation}
E_c(k)\sim \left\{
\begin{array}{ll}
c_i\chi {\epsilon }^{-1/3}k_{*}^{-1}k^{-2/3} & k/k_{*}\ll 1 \\
c_o\chi _o^{-1/3}\chi k_{*}k^{-2} & k/k_{*}\gg 1
\end{array}
\right.   \label{fif}
\end{equation}
where the first form is appropriate for the inverse energy cascade and the
second form is in a form appropriate for the enstrophy cascade. It should be
noted that $k_{*}$ (or $L_{*}$) does not disappear under either asymptotic
limit. In fact as the functional form (\ref{thi}) implies it would be
impossible to eliminate this parameter entirely in both limits.

The spectra obtained above covers a wide range. The range $k/k_{*}<1$
extends to the 10,000 {\it km} wavelength limit imposed by dynamics\cite{Rhi}%
$^{\text{,}}$\cite{Ped} and the $k/k_{*}>1$ extends down to wavelengths of
the order of 100{\it km}, the resolution of the data. Smith and Yakhot \cite
{SmYa} have recently suggested that an inverse cascade for $E\left( k\right)
$ exists for a range of wavelengths greater than 10 {\it km }(but see
Schoeberl and Bacmeister\cite{Obs}). This is based on the assertion that
cumulus cloud activity acts as an energy source. We see no evidence for this
but do not regard this as a contradiction since the resolution of the
satellite data is greater than 100 {\it km}. It is of course vexing that
much the theory and numerical experiments (also\cite{LoCh}) discussed
earlier do not agree with the observed satellite analysis which we present
above. Only the Saffman-Burgers spectrum shows agreement. To test this
further we have looked at the 'discontinuity' patterns of the data and find
that $\left| \bigtriangledown c\right| ^2$ shows filamentous one dimensional
patterns .\cite{Man}A possible explanation for such strand-like patterns has
been discussed recently. Both satellite observations and computer
simulations show the presence of tongues of stratospheric air extending from
the tropics to mid-latitudes. These result from the breaking of Rossby waves
at the edge of the polar vortices.\cite{Nat}

Acknowledgment: The authors are grateful to P. K. Bhartia, B. W. Knight, V.
Yakhot and G. Falkovich for helpful conversations. This work was supported
under a grant from NASA-Goddard (NAG 5-2336).


\begin{references}
\bibitem{Ped}  J.Pedlovsky, {\it Geophysical Fluid Dynamics}, Springer
Verlag (1979)

\bibitem{Phil}  N. A. Phillips, Rev. Geophys. {\bf 1}, 123-76 (1963)

\bibitem{Ssk}  M.R.Schoeberl, R.S.Stolarski and A.Kreuger, Geo.Res.Lett.
{\bf 16}, 377 (1989)

\bibitem{Petal}  M.H. Proffitt et al, Nature {\bf 347}, 31 (1990)

\bibitem{Toms}  Total Ozone Mapping Spectrometer (TOMS) Data, 1978--1993 (P.
Grimares and R. McPeters, eds) NASA Goddard Space Flight Center (1993).

\bibitem{Chem}  R.P. Wayne, {\it Chemistry of Atmospheres}, Oxford (1991)

\bibitem{Sre}  K.R. Sreenivasan, Proc. Roy. Soc. London A {\bf 434}, 165
(1991).

\bibitem{Ko}  A.N.Kolmogorov, Dokl. Akad. Nauk. SSSR {\bf 32}, 19-21 (1941)

\bibitem{ObCor}  A.M. Obukhov, Ize. Acad. Nauk SSSR Geogr. Geofz {\bf 13},
58--69 (1949); S. Corrsin Appl. Phys. {\bf 22}, 469 (1954); A.S. Monin and
A. Yaglom, {\it Statistical Fluid Mechanics Vol. 2}, MIT (1971).

\bibitem{Bat}  G.K. Batchelor, J. Flu. Mech. {\bf 5}, 113 (1958)

\bibitem{Kr1}  R.H. Kraichnan, Phys. Flu. {\bf 11}, 945 (1968)

\bibitem{Sig}  M. Holzer and E.D, Siggia, Phys.Flu. {\bf 6}, 1820 (1994)

\bibitem{Anom}  A. Polyakov, Nucl. Phys {\bf B}(FS) {\bf 396}, 367 (1993);
R.H.. Kraichnan, Phys. Rev. Lett. {\bf 72}, 1016 (1994)

\bibitem{KrBa}  R.H. Kraichnan, Phys. Flu. {\bf 10}, 1417 (1967); G.K.
Batchelor, Phys. Flu. {\bf 125}, II--233 (1969).

\bibitem{Fal1}  G. Falkovich and V. Lebedev, Phys. Rev. E, {\bf 49 }, R1800
(1994)

\bibitem{Comp}  J.C. McWilliams, J. Flu. Mech. {\bf 146}, 21 (1984); R.
Benzi, S. Patarnello, and P. Santangelo, Europhys. Lett. {\bf 3}, 811
(1987); J. Phys. A. {\bf 21}, 1221 (1988); G.F. Carnevale, J.C. McWilliams,
Y. Pomeau, J.B. Weiss, and W.R. Young, Phys. Rev. Lett. {\bf 66}, 2735
(1991); U. Frisch and P. Sulem, Phys. Fluids {\bf 27}, 1921 (1984); L. Smith
and V. Yakhot, Phys. Rev. Lett. {\bf 71}, 352 (1993).

\bibitem{Bo}  V. Borue, Phys. Rev. Lett. {\bf 72}, 1475 (1994).

\bibitem{Obs}  M. R. Schoeberl and J. T. Bacmeister, NATO ASI Series {\bf 18}%
, 135 (1992); D.K. Lilly and E.L. Peterson, Tellus {\bf 35A}, 379 (1983); G.
D. Nastrom, K. S.\ Gage and W. H. Jasperson, Nature {\bf 310}, 36-38(1984).

\bibitem{Fal2}  G.Falkovich and S. Medvedev, Europhys. Lett. {\bf 19}, 279
(1992).

\bibitem{Saf}  P.G. Saffman, Stud.Appl. Math., {\bf 50, }377-383 (1971).

\bibitem{Bur}  J.M. Burgers, Adv. Appl. Mech. {\bf 1}, 171 (1948).

\bibitem{Pie}  R.T. Pierrehumbert, Geophys. Astrophys. Flu. Dyn. {\bf 58 }%
, 285 (1991); R.T. Pierrehumbert, J. Atmos. Sci. {\bf 50 }2462 (1993)

\bibitem{KL}  L. Sirovich and R. Everson, International Journal of
Supercomputer Applications, {\bf 6}(1), 50--68 (1992); G. W. Stewart, SIAM
Rev. {\bf 35}, (1993); G. Berkooz et al, Ann. Rev. Flu. Mech. {\bf 25}
(1993).

\bibitem{Qam}  L. Sirovich, Quarterly of Applied Mathematics, Volume XLV,
Number 3, 561--590 (1987).

\bibitem{Gap}  R. Everson and L. Sirovich (submitted); D. Manin, R. Everson,
B. Knight and L. Sirovich (in preparation).

\bibitem{Man}  D. Manin, R. Everson and L. Sirovich (in preparation).

\bibitem{Prep}  L. Sirovich, R. Everson, B. Knight and D. Manin (in
preparation).

\bibitem{KnSi}  B. Knight and L. Sirovich, Phys. Rev. Lett. {\bf 65}, 1356
(1990).

\bibitem{Hou}  J.T. Houghton, {\it The Physics of Atmospheres}, Cambridge U.
Press (1977)

\bibitem{S85}  L. Sirovich, Phys. Flu. {\bf 28}, 2723 (1985).

\bibitem{Rhi}  P.B.Rhines, J. Flu. Mech. {\bf 69}, 417 (1975).

\bibitem{SmYa}  L.M.Smith and V.Yakhot, J. Flu. Mech. (1994).

\bibitem{LoCh}  E.N. Lorenz, Tellus {\bf 21}, 289 (1969); J.G. Charney, J.
Atmos. Sci. {\bf 28}, 1087 (1971).

\bibitem{Nat}  A. Plumb, Nature {\bf 365 }, 489 (1993); W.J. Randal et al,
Nature {\bf 365 }533 (1993); D. Waugh, Nature {\bf 365}, 535 (1993).
\end{references}
\end{document}